\documentclass[preprint]{emulateapj-rtx4}

\slugcomment{To appear in the AJ}

\usepackage{rotating}
\usepackage{txfonts} 
\usepackage{epsfig}
 \usepackage{graphicx}
\usepackage{natbib} 
\bibpunct{(}{)}{;}{a}{}{,}

\newcommand{\kms}{{\km\second^{-1}}}

\newcommand{\km}{\,{\textnormal{km}}}
\newcommand{\second}{\,{\textnormal{s}}}

\newcommand{\dechms}[4]{$#1^{\rm h}#2^{\rm m}#3\mbox{$^{\rm s}\mskip-7.6mu.\,$}#4$}

\newcommand{\decdms}[4]{$+#1^{\circ}#2'#3\mbox{$''\mskip-7.6mu.\,$}#4$}

\begin{document}

\title{Searching for the main powering outflow objects in Cepheus A East}

\author{Luis A. Zapata\altaffilmark{1}, Manuel Fern\'andez-L\'opez\altaffilmark{2,3}, 
Salvador Curiel\altaffilmark{2}, Nimesh Patel\altaffilmark{4}, and Luis F. Rodr\'\i guez\altaffilmark{1}}
\altaffiltext{1}{Centro de Radioastronom{\'\i}a y Astrof{\'\i}sica,
  Universidad Nacional Aut\'onoma de Mexico, Morelia 58090, Mexico}
\altaffiltext{2}{Instituto de Astronom\'\i a, Universidad Nacional Aut\'onoma de Mexico (UNAM), 
  Apartado Postal 70-264, 04510 Mexico, DF, Mexico}
\altaffiltext{3}{Astronomy Department, University of Illinois, 1002 West Green Street, Urbana, IL 61801, USA}
\altaffiltext{4}{Harvard-Smithsonian Center for Astrophysics, 60 Garden Street, 
Cambridge, MA 02138, USA}

\begin{abstract}
We present (sub)millimeter line and continuum observations in a mosaicing
mode of the massive star forming region Cepheus A East made with the
Submillimeter Array (SMA).  Our mosaic covers a total area of about
3$'$  $\times$ 12$'$ centered in the HW 2/3 region.  For the first time, 
this observational study encloses a high angular resolution ($\sim$ 3$''$) 
together with a large scale mapping of Cepheus A East. 
We report compact and high velocity $^{12}$CO(2-1) emission associated 
with the multiple east-west bright H$_2$ condensations present in the region.  
Blueshifted and redshifted gas emission is found towards the east as well as 
west of HW 2/3.  The observations suggest the presence of multiple 
large-scale east-west outflows that seems to be powered at smaller scales by
radio sources associated with the young stars HW2, HW3c and HW3d.  
A kinematical study of part of the data suggests that the molecular outflow 
powered by HW2  is precesing with time as recently reported.  
Our data reveal five periodic ejections of material separated
approximately every 10$^\circ$ as projected in the plane of the sky.   
The most recent ejections appear
to move toward the plane of the sky.  An energetic explosive event as
the one that occurred in Orion BN/KL or DR21 does not explain the kinematics,
 and the dynamical times of the multiple ejections found here. The
continuum observations only revealed a strong millimeter source
associated with the HW 2/3 region. High angular resolution observations 
allow us to resolve this extended dusty object in only two compact sources (with spatial sizes of
approximately 300 AU) associated with HW2 and HW3c. Finally, the
bright optical/X-Ray HH 168 -- GDD37 object might be produced by strong 
shocks related with the outflow from HW3c.
\end{abstract}


\section{Introduction}

Located at a distance of 700 pc \citep{mosca2009,sergio2011}, Cepheus A East is 
one of the nearest high-mass star formation regions. For this reason, Cepheus A East
like Orion, is considered an ideal laboratory for the study of the physical processes related with the 
formation of the massive stars. This region has a total 
bolometric luminosity of about 2.5 $\times$ 10$^4$ L$_\odot$ \citep{eva1981}. 
Half of this luminosity is attributed to the HW2 object, the brightest radio continuum 
source in the field, which is considered to be a B0.5 spectral type protostar \citep{rod1994}.
 
Cepheus A East contains a massive bipolar molecular outflow aligned
primarily east-west \citep{Rod1980}, but with additional components
aligned northeast-southwest \citep{Bally1991,chema1993,gom1999}.  
The central 2$'$ region contains high velocity as well as more
compact extremely high velocity $^{12}$CO components with radial
velocities ranging from $+$50 to $-$70 km s$^{-1}$ relative to the CO
centroid \citep{nara1996}.  The axis of the extremely high velocity
outflow is rotated roughly 40$^\circ$ clockwise relative to the high 
velocity outflow on the plane of the sky. 
The smaller spatial extent together with the higher
velocity suggests that the extremely high velocity flow traces a younger outflow
component. At low velocities, there are additional blueshifted and
redshifted components centered on HW2 that are oriented
northeast-southwest.  

This outflow complex also contains several
Herbig-Haro objects, including the extremely bright HH 168 or GGD 37 located
about 100$''$ due west of HW2, and several fainter bow shocks located
to the east \citep{har2000}.  Fainter HH objects (HH 169 and 174) are
located in the eastern lobe \citep{har2000}.  \citet{pra2005} reported several 
distinct soft X-ray sources within the GGD 37 flow, one near the west, 
and one at the far end of the flow. They concluded that  GGD 37 has a luminosity in X-Rays of
L$_X$ = 3 $\times$ 10$^{30}$ erg s$^{-1}$, and inferred a shock velocity 
of 620 km s$^{-1}$ for this object.

The complexity of the outflows, their multiple orientations and
velocities, and the unclear morphology of some of the shock features
make Cepheus A East challenging to interpret \citep{cun2009}.
However, the study of the kinematics, and morphology of the outflows 
ejected from the high-mass protostars located in Cepheus A East will allow us 
to understand better how the massive stars form in a clustered region. 

Recent sensitive near-infrared H$_2$ line observations, and thermal
infrared observations of this nearby massive star-forming region
indicate that the massive young stellar object HW2 drives a
pulsed, precessing jet that has changed its orientation by about
45$^\circ$ in roughly 10$^4$ years \citep{cun2009}. This outflow thus
appears to be responsible for most of the outflowing activity observed
in Cepheus A.  The precessing outflow is possibly feed by the
massive and large dusty disk located in HW2 \citep{patel2005,chema2007,Ji2009}, and is
associated with a bright and powerful radio thermal jet \citep{cur2006,rod2005,rod1994}.  
\citet{cun2009} proposed that the close passages of
an intermediate mass companion through or near the circumstellar disk
result in periods of enhanced accretion and mass loss, as well as
forced precession of the disk. However, the presence of faint HH objects and molecular
outflows with different orientations close to the HW2 object and
towards the HH 168 object suggest that this precessing-jet model may not be 
the only mechanism working at this region to produce the
outflowing activity.

At the moment, there have been only a few successful attempts to study
the kinematics of this region after the pioneering single dish $^{12}$CO
single dish observations made by \citet{Rod1980}. Some examples include the
extremely high angular resolution VLBI studies of water masers
\citep{chema2011} which sample the kinematics only sparsely, due to
the maser's high selectivity, the proper motions measured in the radio jets present in this 
region \citep{rod2005,cur2006}, and  the large scale SiO and HCO$^+$ maps
from \citet{gom1999}. 

In the innermost parts of Cepheus A East there are multiple compact radio sources revealed 
mainly by sensitive Very Large Array (VLA) observations.  
For example, HW3 c/d are radio sources associated with a cluster of H$_2$O masers
located at the core of this region.  They are located 5$''$ south
of HW2 and close to the western end of a nearly continuous chain of
radio sources along the southern rim of the Cepheus A East radio
source complex.

\citet{hu1984} suggested that the bulk of the Cepheus A's luminosity
($\sim$ 10$^4$ L$_\odot$) likely arises from the radio sources HW2 and
HW3 c/d that are associated with bright H$_2$O masers. 
However, no submillimeter continuum emission has been detected 
from HW3d, leading \citet{prepo2007} conclude that HW3c is most likely 
to harbor the second most luminous and massive YSO in the Cepheus A  East core.  
In this interpretation, radio source 3d may trace part of an elongated thermal jet from
HW3c.

In this paper, we present high angular resolution 
(sub)millimeter continuum and line observations of the region Cepheus A
East made with the Submillimeter Array\footnote{The
Submillimeter Array (SMA) is a joint project between the Smithsonian
Astrophysical Observatory and the Academia Sinica Institute of
Astronomy and Astrophysics, and is funded by the Smithsonian
Institution and the Academia Sinica.} (SMA).  This is the first $^{12}$CO(2-1) and submillimeter continuum 
study of Cepheus A East that contains a high angular resolution ($\sim$ 3$''$) together 
with a large scale mapping (3$'$  $\times$ 12$'$). In Section 2, we
discuss the observations made in this study. In Section 3, we present the data. 
In Section 4 we discuss the data, and in Section 5, we give the main conclusions.

\section{Observations}

\subsection{Millimeter}

The observations were made with the SMA on 2010 August 17th and 2010 October 17th, 
in its subcompact and compact configurations, respectively.  
The phase reference center for the observations was at
$\alpha_{J2000.0}$ = \dechms{22}{55}{51}{44}, $\delta_{J2000.0}$ = 
\decdms{62}{01}{51}{3}. The frequency was centered at 220.53797 GHz in the 
Lower Sideband (LSB), while the Upper Sideband (USB) was centered at 230.53797 GHz. 
The primary beam of the SMA at around 230 GHz has a FWHM of about
50$''$.  We used the mosaicing mode with half-power point spacing between 
field centers and covered the entire Cepheus A East outflow region as far as it was 
mapped in H$_2$ by \citet{cun2009}, see Figure 1. We concatenated the two 
data sets using the task in MIRIAD called "{\it uvcat}''. The two 
different observations were identical, and only the antenna configuration
of the SMA changed. The greatest angular size source that can be imaged 
on these observations is approximately 25$''$ at a 10\% level \citep{wil1994} . 
The astrometric errors of these observations are about of 1$''$.

The SMA digital correlator was configured in 24 spectral windows
(``chunks'') of 104 MHz each, with 256 channels distributed over each
spectral window, providing a resolution of 0.4062 MHz ($\sim$ 0.5 km
s$^{-1}$) per channel. However, in this study, we smoothed the spectral 
resolution to  about 1.5 km s$^{-1}$, because of the large width of the $^{12}$CO line 
toward the Cepheus A East region \citep{Rod1980}. The total bandwidth
for both SMA observations is 8 GHz. 

The zenith opacity ($\tau_{230 GHz}$), measured with the NRAO tipping
radiometer located at the Caltech Submillimeter Observatory, was from
0.2 to 0.05, indicating good weather conditions during the
observations. Observations of Uranus provided the absolute scale for
the flux density calibration.  
The gain phase calibrators were BLLAC, J2202+422, and J0014+612.  
The bandpass calibrator was the quasar 3C89.
The uncertainty in the flux scale is estimated
to be between 15 and 20$\%$, based on the SMA 
monitoring of quasars. Further technical descriptions of the SMA, 
and its calibration schemes can found in \citet{Hoetal2004}.

The data were calibrated using the IDL superset MIR, originally
developed for the OVRO \citep{Scovilleetal1993} and adapted for the
SMA\footnote{The MIR-IDL cookbook by C. Qi can be found at
  http://cfa-www.harvard.edu/$\sim$cqi/mircook.html}. The calibrated
data were imaged and analyzed in the standard manner using the MIRIAD, AIPS,
and KARMA \citep{goo96} softwares.  We set the ROBUST parameter of the task INVERT
to $+$2 to obtain a slightly better sensitivity sacrificing some
angular resolution. After concatenated the two 
data sets the resulting rms noise for the line images was around
150 mJy beam$^{-1}$ for each velocity channel (2 km s$^{-1}$) and 15 mJy for the continuum 
at an angular resolution of $3\rlap.{''}9$ $\times$ $3\rlap.{''}1$ with a P.A. =
$-$$11.6^\circ$. The $^{12}$CO(2-1) line was detected at a frequency of about 230.5 GHz.
Many more lines were detected in our spectral bands, however, most of them are
associated with the compact object HW2 and do not with the multiple outflows emanating 
from this region. Thus, on this study, we only concentrated on the spatial distribution of the CO line
emission. Moreover, even the compact emission from the SiO that is a natural outflow 
tracer is located only near HW2.

Here, we will assume a systemic velocity for the molecular cloud of $-$10 km s$^{-1}$ as
reported in \citet{pin2005}.

\subsection{Submillimeter}

The observations were obtained from the SMA archive and were collected with 7 antennas of 
the SMA on 2005 November 30 in its very extended configuration.  The receivers were tuned to a
frequency of  341.0715 GHz in the USB, while the LSB was centered on
331.0715 GHz. The phase reference center for the observation was
$\alpha_{J2000.0}$ = \dechms{22}{56}{17}{98}, $\delta_{J2000.0}$ = 
\decdms{62}{01}{47}{9}. The total bandwidth for this SMA observation is 4 GHz. 

The zenith opacity ($\tau_{230 GHz}$) was about 0.06, indicating excellent weather conditions.  
Observations of Ceres provided 
the absolute scale for the flux density calibration.  
The uncertainty in the flux scale is estimated
to be again between 15 and 20$\%$.
The gain phase calibrator was the quasar BLLAC.
The bandpass for this observation was the quasar 3C454.3.
The continuum rms noise was
around 15 mJy beam$^{-1}$ at an
angular resolution of $0\rlap.{''}74$ $\times$ $0\rlap.{''}73$ with a
P.A. = $32.7^\circ$. The data was self-calibrated in phase and amplitude.

\begin{figure*}[ht]
\includegraphics[scale=0.35, angle=0]{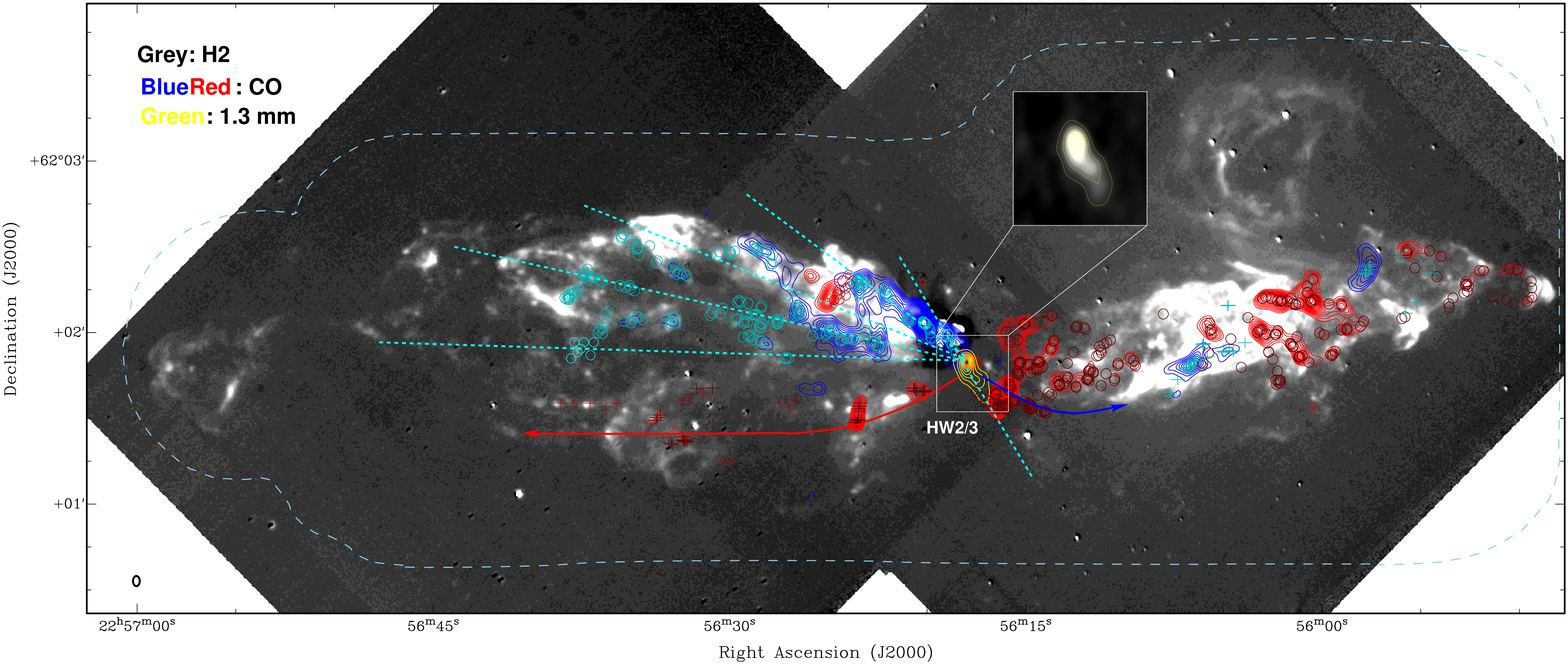}
\caption{\scriptsize Integrated intensity contour map of the
$^{12}$CO(2-1) emission overlaid on the H$_2$ image presented in
\citet{cun2009}.  The redshifted and blueshifted emission are
represented by red and blue contours, respectively.  The blue and red
contours start from 10\% to 90\% with steps of 3\% of the peak of the
molecular emission; the peak of the line map is 85 Jy beam$^{-1}$ km
s$^{-1}$ for the blueshifted gas and 75 Jy beam$^{-1}$ km
s$^{-1}$ for the redshifted gas.  The range for the integrated redshifted emission 
is from $+$5 to $+$55 $\kms$, while the range for the integrated blueshifted 
emission is from $-$65 to $-$14 $\kms$.
The systemic velocity assumed here is  $-$10 km s$^{-1}$.
The yellow contours represent the 1.3 mm continuum emission.
The yellow contours are from $+$10\% to 90\% with steps of 10\% of the
peak of the emission; the peak of the continuum map is 1.0 Jy
beam$^{-1}$.  Additionally, the blue/red circles and crosses mark
the position of the $^{12}$CO emission condensations found at different velocity channels. 
The $^{12}$CO emission close to the systemic velocity was poorly
sampled with the SMA and is not presented in this image.  The open
circles trace emission likely arising from HW2, while the crosses trace the
emission arising from the HW3 region (see text).  The continued and dashed
lines mark the orientations of the outflows emanating from HW2 and HW3, respectively. The
blue dashed line at the edge of the H$_2$ image marks the area that
our SMA mosaic covers. The synthesized beam of the line and continuum 
observations is shown in the lower left corner.  The inset in the image shows a zoom toward
the continuum millimeter source reported here. }
\label{fig1}
\end{figure*}

\begin{figure*}[ht]
\includegraphics[scale=0.33, angle=0]{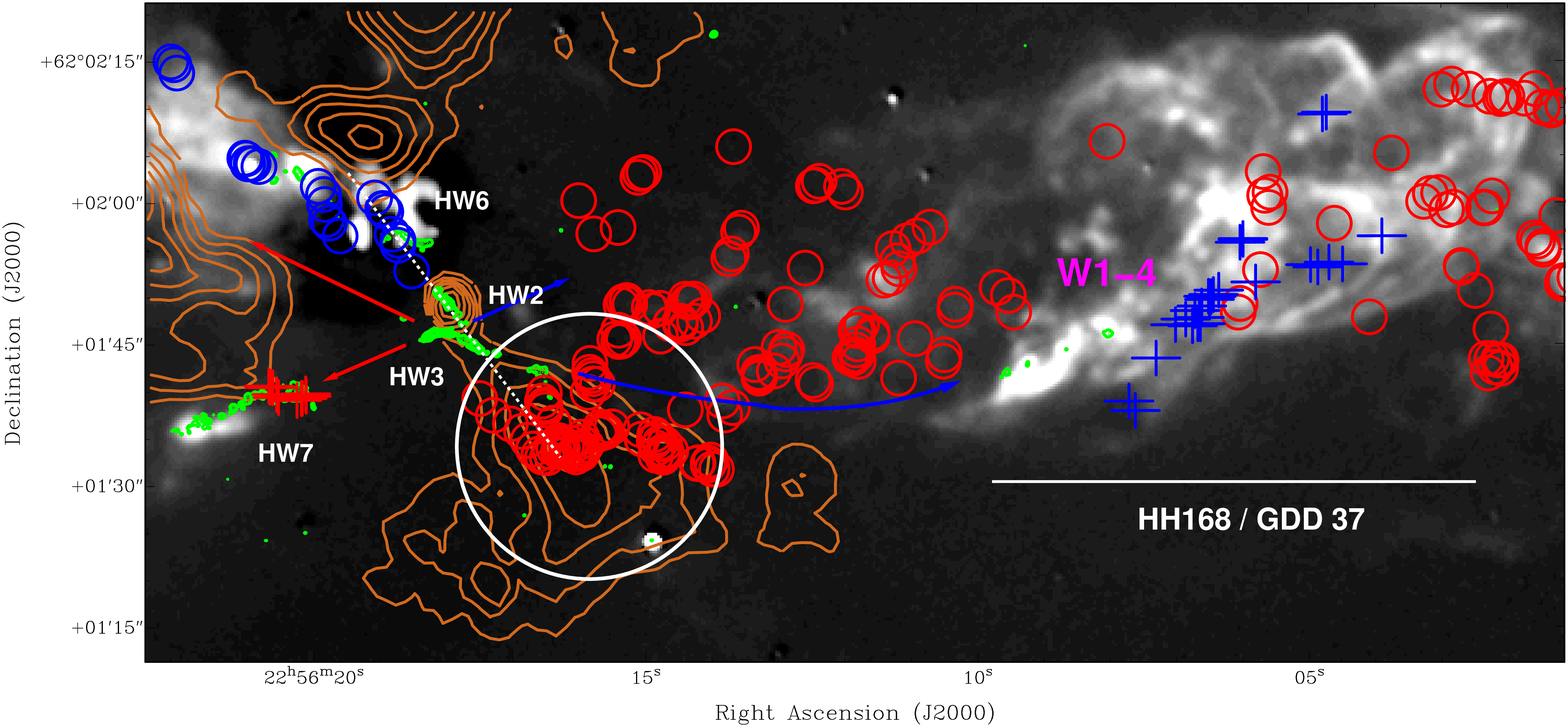}
\caption{\scriptsize Zoom into the HW2 / HW3, and the HH 168 regions. 
As in the case of Figure 1, 
the blue and red open circles/crosses mark the
position of the $^{12}$CO emission peaks at different velocity channels, 
and are overlaid on an H$_2$ image.  The green contours trace the VLA radio
emission \citep{cur2006,rod1994}. The green contours are from $+$5\%
to 90\% with steps of 1\% of the peak of the emission; the peak of the
continuum map is 4.0 mJy beam$^{-1}$.  The brown contours trace the
NH$_3$(1,1) emission \citep{chema1993}.  The physical and observational
parameters of these NH$_3$ condensations are given in \citet{chema1993}.
The brown contours start from
15\% to 90\% with steps of 10\% the peak of the molecular emission;
the peak of the line map is 257 Jy beam$^{-1}$ km s$^{-1}$. The white 
circle marks the position of the "shocked-zone" described in the text.}
\label{fig2}
\end{figure*}

\clearpage

\begin{deluxetable*}{l c c c c c c c c}
\tabletypesize{\scriptsize}
\tablecaption{Parameters of the SMA (sub)millimeter sources}
\tablehead{                        
\colhead{}                        &
\multicolumn{2}{c}{Position} &
\colhead{}                              &
\multicolumn{3}{c}{Deconvolved size$^a$} &        \\
\colhead{}   &
\colhead{$\alpha_{2000}$}          &
\colhead{$\delta_{2000}$}           &
\colhead{Flux Density }       &                            
\colhead{Maj.}  &
\colhead{Min.}  &
\colhead{P.A.}  &
 \colhead{Mass}  & \\
\colhead{Source}                              &
\colhead{(h m s) }                     &
\colhead{($^\circ$ $^{\prime}$  $^{\prime\prime}$)}              &
\colhead{(mJy)}  & 
\colhead{($^{\prime\prime}$)}  &
\colhead{($^{\prime\prime}$)}  &
\colhead{($^\circ$)} &
\colhead{($M_\odot$)} &
}
\startdata

HW2/3   &    22 56 17.999 & $+$62 01 48.82 &   2.1 $\pm$ 0.3    &  5.3 $\pm$ 0.5  &  0.7 $\pm$ 0.2 & $-$13.1 & 5 \\
HW2      &    22 56 17.978 & $+$62 01 49.39 &    0.9 $\pm$ 0.2  &  0.77 $\pm$ 0.05  &  0.3 $\pm$ 0.1 & $-$152 & 0.5 \\
HW3c    &    22 56 17.943 & $+$62 01 46.04 &    0.4 $\pm$ 0.2   &    --    &     --    &  -- & 0.3 \\
\enddata
\tablecomments{
                (a): These values were obtained from the task IMFIT of MIRIAD.}
\end{deluxetable*}

\section{Results}

\subsection{Millimeter continuum emission}

In Figure 1, we show the resulting 1.3 mm. continuum image from the
Cepheus A East from our SMA observations.   
The continuum emission is represented by yellow contours in this Figure.
The mosaic covered a total area of
about 3$'$ $\times$ 12$'$ centered at the HW 2/3 region. We only detected a
strong and compact source associated with HW 2/3. The source has additionally a
small tail that extends to the southwest by a few arcseconds.  A similar
morphology was already reported by \citet{prepo2007}.  
The physical and observational parameters of this source are
presented in Table 1. The compact millimeter
source is coincident with a strong and compact condensation of NH$_3$
associated with HW2, see Figure 2. The physical and observational values for this
compact NH$_3$ condensation are given in \citet{chema1993}.

Following to \citet{zap2012} and assuming a
dust temperature of 50 K,  a distance of 700 pc, a $\beta$ = 1.5, optically thin and isothermal 
dust emission, and a gas-to-dust ratio of 100, we found a dust-mass detection 4$\sigma$ limit
of 0.5 M$_\odot$. This suggests that the presence of a cluster of young very low-mass stars
might be present in Cepheus A East.  

\subsection{Submillimeter continuum emission}

In Figures 3 and 4, we show the submillimeter continuum maps obtained
with the SMA.  We found submillimeter compact
continuum emission associated with only two objects, HW2 and HW3c. 
This is in agreement with the submillimeter maps
presented in \citet{prepo2007}. However, in their maps they found a
very faint source called SMA4 that is not detected here at a 4$\sigma$
level. Both maps, the presented here and that obtained by \citet{prepo2007},
have similar rms-noises so that we believe that such differences could come
from the different {\it uv-coverage} obtained in the observations. The observational
and physical parameters of these sources are given in Table 1.
The millimeter source associated with HW3c is
not resolved.  The values for the fluxes obtained here from both
sources (HW2 and HW3) are also in good agreement with the values
obtained in \citet{prepo2007}. To obtain the physical dimensions and flux densities 
of the continuum sources we used the task {\it imfit} of the MIRIAD. This routine
uses gaussian fitting to derive these parameters.

\begin{figure}[ht]
\begin{center}
\includegraphics[scale=0.33]{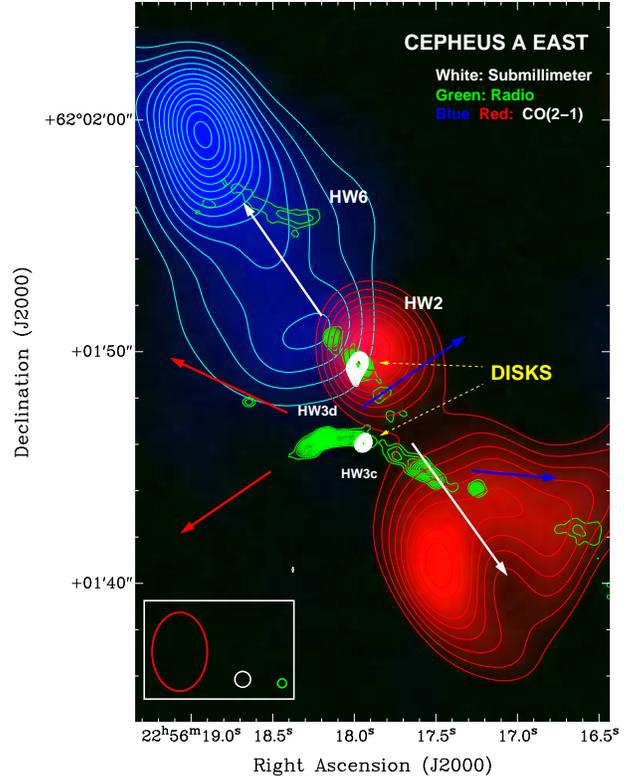}
\caption{\scriptsize Integrated intensity color and contour maps of
the $^{12}$CO(2-1) emission from the HW 2/3 region overlaid in contours
with the SMA 890 $\mu$m continuum emission (white) and the VLA 6 cm
continuum emission (green). The blue and red contours are from 30\% to
90\% with steps of 10\% of the peak of the line emission; the peak of
the $^{12}$CO(2-1) emission is 24 Jy beam$^{-1}$ km s$^{-1}$ for the 
blueshifted gas and 18 Jy beam$^{-1}$ km s$^{-1}$ for the redshifted gas.  The
integration range for the blueshifted velocities was $-$22 km s$^{-1}$
to $-$19 km s$^{-1}$, while for the redshifted velocities was $-$4 km
s$^{-1}$ to $-$1 km s$^{-1}$.   The systemic velocity assumed here is 
$-$10 km s$^{-1}$. The white contours are from 40\% to
90\% with steps of 5\% of the peak of the emission; the peak of the
890 $\mu$m map is 420 mJy beam$^{-1}$.  The green contours are from
10\% to 90\% with steps of 3\% of the peak of the emission; the peak
at 6 cm is 4.0 mJy beam$^{-1}$.  The synthesized beams of the line and continuum 
observations are shown in the bottom left corner. The white, blue, and red
arrows represent the orientation of the multiple outflows emanating from this region.
The molecular emission extending outside of the figure is associated with multiple 
east-west outflows.}
\label{fig3}
\end{center}
\end{figure}   

\subsection{$^{12}$CO(2-1) line emission}

In Figure 1, we show the main results of the line emission analysis. In this
image, we present the integrated intensity of the line emission
(moment zero map) together with the position of each gas compact condensation
found in our spectral channel maps.  This figure is made in this way because the
dominant parts of the outflow do not allow us to see clearly fainter molecular gas
in our moment zero map. This effect is clearer far from
the center of the image, where the $^{12}$CO line emission is mainly extended and faint. 
The blue and red contours represent the
$^{12}$CO gas emission that is approaching and receding from us,
respectively.  Receding (redshifted) $^{12}$CO features show radial
velocities down to 65 $\kms$ while the approaching molecular $^{12}$CO 
gas has values up to $-$55 $\kms$.
The moment zero map is overlaid with the H$_2$ emission, which is
related with shocked gas and is presented in \citet{cun2009}.  It is
evident that there is a good alignment between the $^{12}$CO and H$_2$
emission. In both sides of the outflow redshifted and
blueshifted gas is found, suggesting the presence of multiple outflows with
different orientations emanating from the HW 2/3 region.  Figure 1 only
displays the most prominent $^{12}$CO emission features detected
outside of the velocity window $-$13 to $-$4 $\kms$.  Within this window
the radiation stems predominantly from the ambient cloud and is
spatially extended, thus cannot be properly reconstructed by the SMA.
On the other hand, the emission at high velocities is generally compact \citep[e.g.][]{rod1999}, 
and is properly reconstructed by the SMA, see for example the outflow located 
in Orion-KL \citep{peng2012}.  Here, we take as compact emission, 
$^{12}$CO molecular structures with sizes of about two or three synthesized beams. 
    
In Figures 1 and 2, we have distinguished the molecular material
ejected by HW2 and HW3c/d with blue and red open circles for HW2 and
blue and red crosses for HW3c/d. We are able to do this because of the
different orientations of the molecular outflows. For example, the redshifted 
and blueshifted emission for the outflow from HW2 is mostly located in the west and
east, respectively. The outflow powered by HW3c approximately has its redshifted side
in the northeast and its blueshifted side in the west.

In Figures 3 and 4, we show $^{12}$CO moment zero maps using a
limited velocity window, trying therefore to reveal young molecular gas 
(or gas that is closer to the exciting sources as compared to the most eastern gas) associated with HW2 and HW3.  
Both maps show bipolar collimated outflows emanating from each of these sources.
The molecular outflows appear to emanate from the compact submillimeter and radio
sources associated with HW2, HW3d, and HW3c. At very small scales (about 2000 AU) and in the
middle of the flows, it is found the compact radio emission reported by
\citet{cur2006,rod1994}.  The radio emission associated with HW2 is
related to a powerful radio thermal jet with tangential velocities of
about 500 $\kms$ \citep{cur2006}.  
The nature of the radio emission associated with HW3c/d, on the other hand,
is still unknown. However, their spectral indices and morphologies are
consistent with possible thermal jets with different orientations
\citep{garay1996,chi2012}.  No clear proper motions of the ionized gas are
observed toward these sources \citep{rod2005}. Here, we assume that these
sources are in fact extended thermal jets, however, more observations 
are still needed to confirm this hypothesis.  

\begin{figure*}[ht]
\begin{center}
\includegraphics[scale=0.32, angle=0]{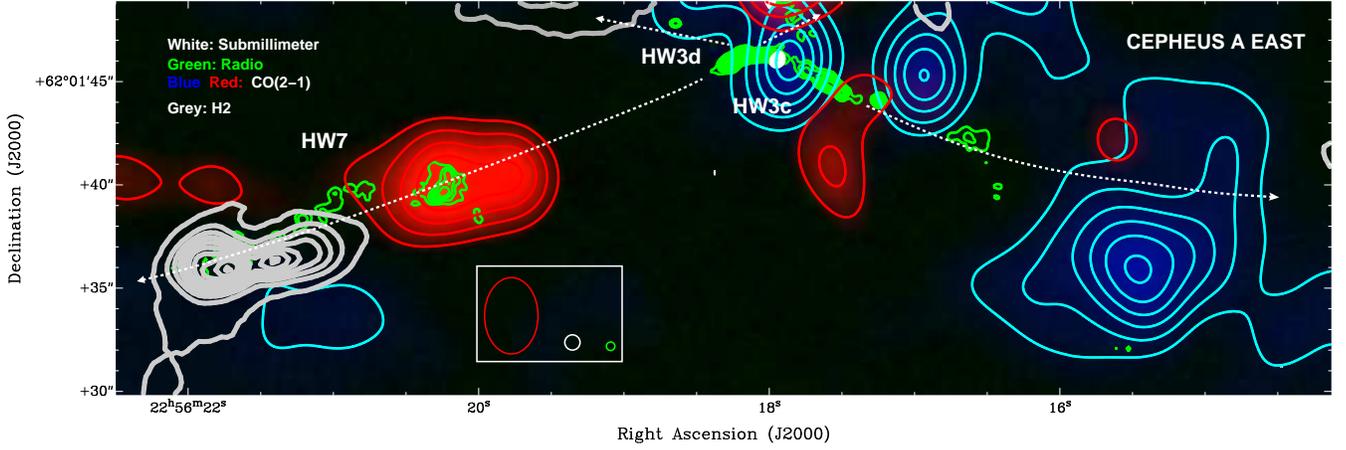}
\caption{\scriptsize Integrated intensity color and contour maps of the $^{12}$CO(2-1) 
emission from the HW3 region overlaid in contours with the SMA 890 $\mu$m continuum 
emission (white), infrared H$_2$ emission (grey) and the VLA 6 cm continuum emission (green). 
The blue and red contours are from 40\% to 90\% with steps of 10\% of the peak of the line emission; 
the peak of the $^{12}$CO(2-1) emission is 17 Jy beam$^{-1}$ km s$^{-1}$ for the redshifted gas,
and 12 Jy beam$^{-1}$ km s$^{-1}$ for the blueshifted gas. 
The integration range for the blueshifted velocities was $-$17 km s$^{-1}$ to  $-$14 km s$^{-1}$, 
while for the redshifted velocities was $-$7 km s$^{-1}$ to  $-$4 km s$^{-1}$.
The systemic velocity assumed here is $-$10 km s$^{-1}$.
The white contours are from  40\% to 90\% with steps of 5\% of the peak of the emission; 
the peak of the 890 $\mu$m map is 420 mJy beam$^{-1}$. 
The green contours are from 10\% to 90\% with steps of 3\% of the peak of the emission; 
the peak at 6 cm is 4.0 mJy beam$^{-1}$.   The synthesized beams of the line and continuum 
observations are shown in the lower left-middle corner. The white arrows represent the orientation 
of the multiple outflows emanating from this region. }
\label{fig4}
\end{center}
\end{figure*}

For the case of HW2, the
redshifted emission is found in the southwest while blueshifted
emission is in the northeast. For HW3c, the redshifted emission is
found to the east while blueshifted emission is in the west (see Figure 4).  
There is some redshifted emission located in northeast of HW2/3 region (Figure 1),
maybe this is part of this bipolar outflow powered by HW3c. 
For HW3d, we think that the redshifted emission found to the southeast (towards HW7) is related 
with this object, but there is 
no evidence from our observations of blueshifted emission towards the northwest,
probably because the blueshifted outflowing material goes directly outside 
of the molecular cloud as it seems to be case in some others molecular outflows \citep{che1995,zap2010}. 
This picture is consistent with the SE-NW outflow traced by the water proper motions, and 
the orientation of the thermal jet reported by \citet{chi2012} in HW3d.

The small changes observed in the orientation of the different outflows found in Cepheus A 
east can be explained by the outflows undergoing deflections, possibly due to the high density
molecular cloud. A process for the deflection of an outflow has been proposed and 
modeled by \citet{raga1995}. However, again many more observations and theoretical studies 
are still needed to confirm this proposition. 

The$^{12}$CO emission arising from these outflows is also associated with
H$_2$ and radio emission located far from the energizing sources 
({\it e.g.} the zone close to HW7 and HW6). 
This radio emission is probably produced by strong shocks in
dense zones \citep{rod2005}.

The orientation of the outflows reported here are in agreement with
the position of the multiple outflows emanating from this region
proposed by \citet{goe1998}, see their Figure 7.

\section{Discusion}

\subsection{Dusty Envelope and Circumstellar Disks}

Following \citet{zapata2012}, we adopted a value of $\kappa_{\rm 1.3
mm}$ = 0.1 cm$^2$ g$^{-1}$ \citep{oss1994} and assumed optically thin,
isothermal dust emission, a distance to Cepheus A East of 700 pc 
\citep{mosca2009,sergio2011} and a
gas-to-dust ratio of 100, with a dust temperature of 50 K for the
extended millimeter object associated with HW 2/3, deriving a
dust mass of 5 solar masses.  At these wavelengths, 
the emission from the ionized jets is negligible because its emission
depends on frequency as $\nu^{0.6}$ \citep{rey1986}.   
The mass and linear sizes ($\sim$ 2000
AU) suggest that this dusty source (shown in Figure 1) is likely a
large envelope surrounding the compact objects HW2 and HW3. 

For the case of the submillimeter compact sources, we found masses of about
0.5 M$_\odot$. We assumed a dust temperature of 100 K, and a dust opacity 
of $\kappa_{\rm 0.8 mm}$ = 0.15 cm$^2$ g$^{-1}$.
The mass and linear size ($\sim$ 350 AU) of HW2 are
consistent with the values obtained for the circumstellar disk
reported by \citet{patel2005,chema2007,Ji2009}. However, the
circumstellar disk could be not only forming a single massive star
\citep{cur2002,com2007,prepo2007}.  For the case of HW3c, we suggested that the dusty
object mapped here and shown in Figure 3 is also a circumstellar disk
that powers the east-west bipolar outflow shown in Figure 4. A more compressive discussion 
about this hypothesis will give in a future paper (Curiel et al. 2013). 

\subsection{Molecular Outflows}

\subsubsection{The pulsed and precessing outflow from HW2}

We found strong and compact molecular $^{12}$CO line emission associated with the
pulsed and precessing outflow reported by \citet{cun2009}.  The
redshifted emission is found to the west while the blueshifted is
found to the east.  The blueshifted emission is mostly distributed in
compact condensations that follow well the H$_2$ emission. Very close
to HW2 well defined jet-like structures are observed. The redshifted
emission, on the other hand, is distributed regularly forming arcs
with different orientations and sizes.  Most of the molecular arcs are
formed on the tips of the H$_2$ bow shocks.  We found $^{12}$CO
blueshifted and redshifted emission at distances up to 1 pc away from
HW2. We do not detect $^{12}$CO emission from the most eastern H$_2$
bow shock (HH 174). From our data it is not clear whether the object HH 174 was  
ejected from HW2 as suggested by \citet{cun2009}, or maybe this was
produced by HW3c/d long time ago. Most of $^{12}$CO emission detected by
\citet{Rod1980} and \citet{cun2009} using single dishes, seems to be
extended and was likely resolved out in our high resolution maps.
 
From our $^{12}$CO map, presented in Figure 1, it is not easy to
see the eastern ejections formed at different angles of the
precessing outflow driving by HW2 and reported by \citet{cun2009}.  This is only
clear near HW2.  The emission is mostly concentrated in small groups
dispersed across the H$_2$ bow shocks.
In Figure 5, we show a part of the velocity cube obtained here trying to illustrate
how we get the physical information of each compact condensation present in our SMA data.   
If we plot the position angle {\it vs.} the projected distance with respect to the position of HW2 of each $^{12}$CO
condensation found in our spectral channel cube (see Figure 6), one
can distinguish groups of condensations separated approximately every 10$^\circ$.  
This result is in very good agreement
with that reported by Cunningham et al. (2009) and that is shown
in our Figure 1. In a first approximation, the condensations
with small position angles are close to HW2, while the
condensations with large angles are far from HW2. 
This is also the case for the west side of the outflow (see Figure 5).
For example, condensations that have PAs between 40$^\circ$ to 60$^\circ$
are located in a range of 10$''$ to 20$''$ from HW2 (very close of this object), in the other hand,
condensations with PAs about 93$^\circ$ are far from HW2, about  250$''$. 
However, there are some condensations that show a large spread of distances 
in a small range of angles. These condensations have an average angle of
85$^\circ$ and are associated with a single ejection from HW2. 
The small differences between both diagrams (Figures 5 and 6) are probably due to
inhomogeneities of the molecular cloud.  

\begin{figure}[ht]
\centering
\includegraphics[scale=0.23, angle=0]{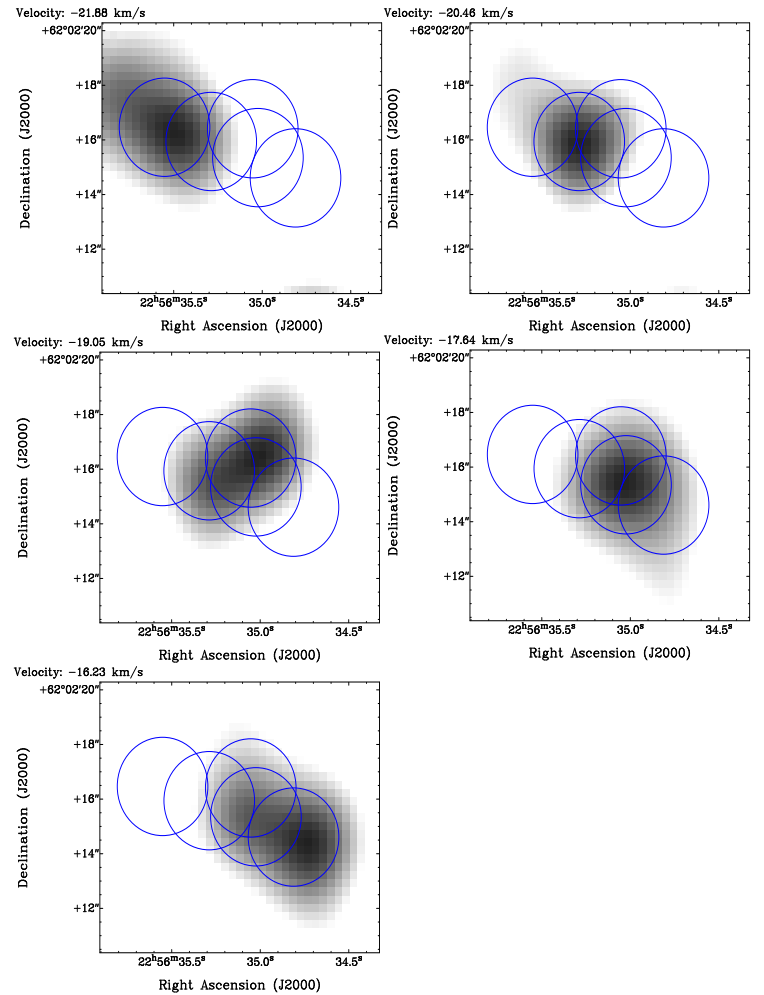}
\caption{\scriptsize A part of the velocity cube for a single blueshifted filament showing how
the physical information of each filament using a gaussian fitting was obtained.  
Note that approximately the size of every molecular condensation is about one beam. The blue
circles mark the position of the condensations.}
\label{fig3}
\end{figure}

\begin{figure}[h!]
\begin{center}
\includegraphics[scale=0.52, angle=0]{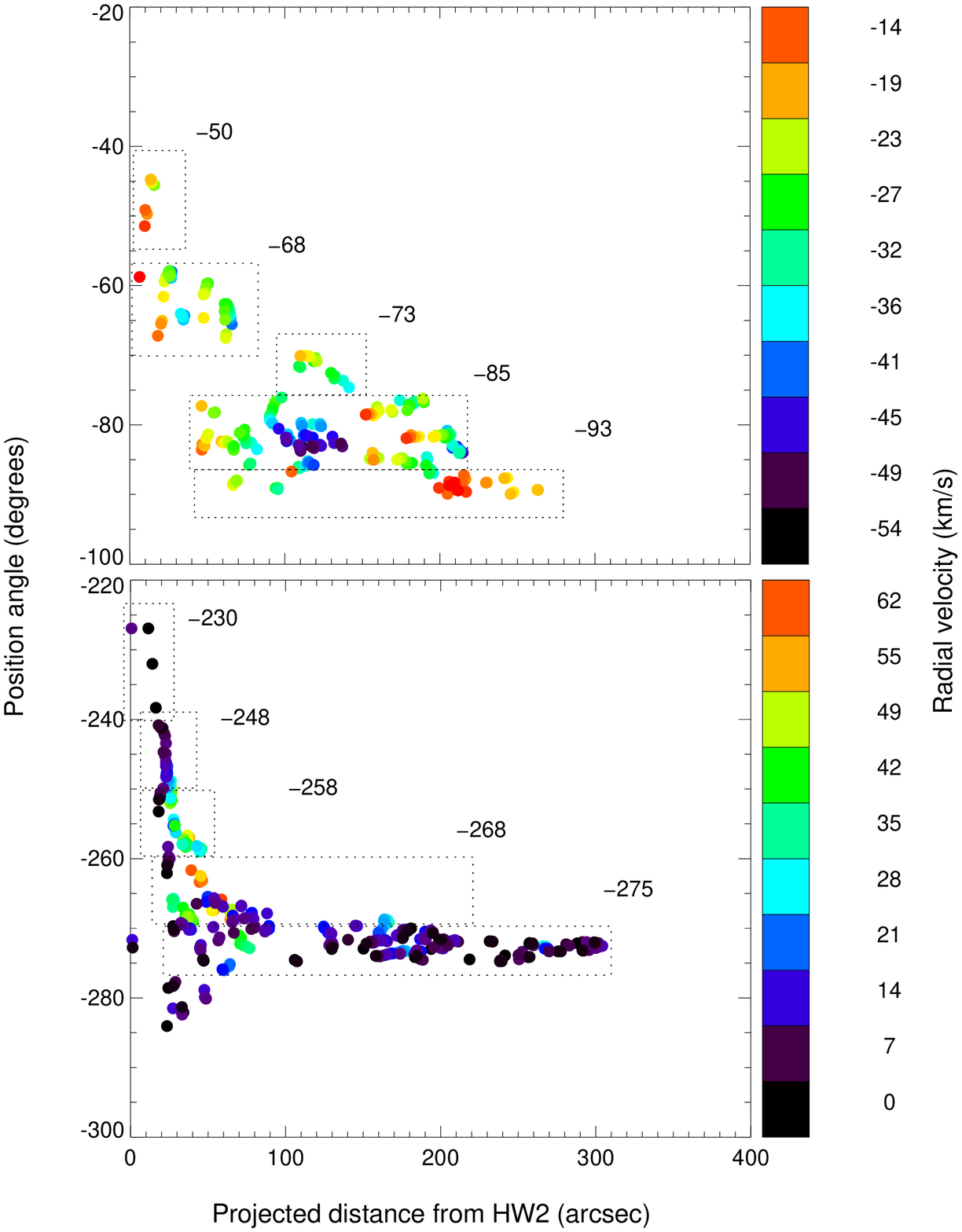}
\caption{\scriptsize Position angle vs. projected distance to HW2
digram for the blueshifted and redshifted $^{12}$CO spots.  The
$^{12}$CO spots are additionally color-coded to their respective
radial velocities.  The dashed boxes show the groups of $^{12}$CO
condensations for different position angles.  }
\label{fig5}
\end{center}
\end{figure}

\begin{figure}[ht]
\begin{center}
\includegraphics[scale=0.48, angle=0]{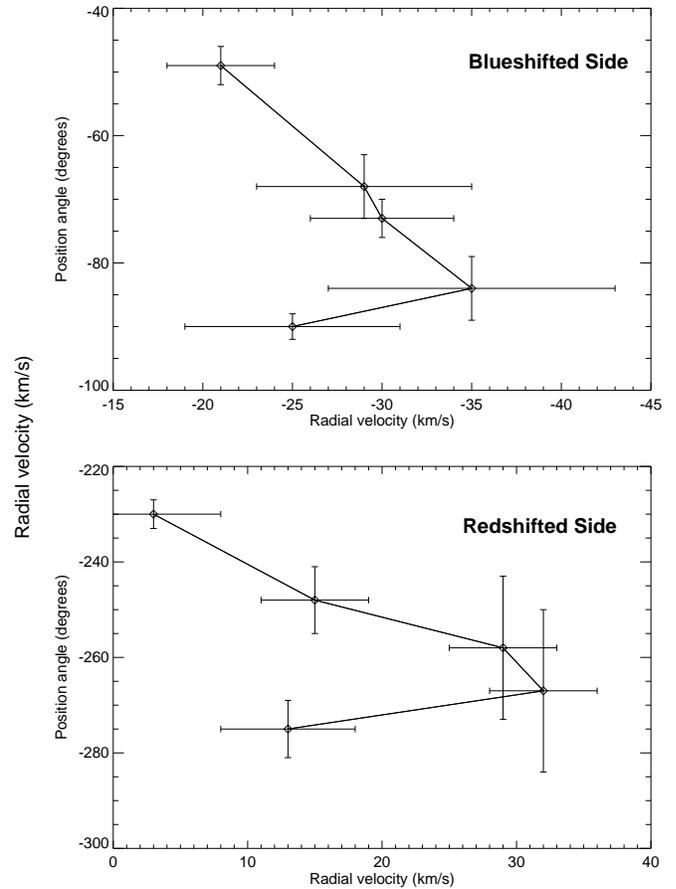}
\caption{\scriptsize Similar plot as the one presented in Figure 5,
but only averaging the $^{12}$CO condensations in different groups of
position angles. The average velocity and position values show
their respective error bars. The error bars are the standard error on the mean.
The edges of the bins
here are an approximation trying to select the molecular condensations in
groups of 10$^\circ$.}
\label{fig6}
\end{center}
\end{figure}

\begin{figure}[h!]
\begin{center}
\includegraphics[scale=0.45, angle=0]{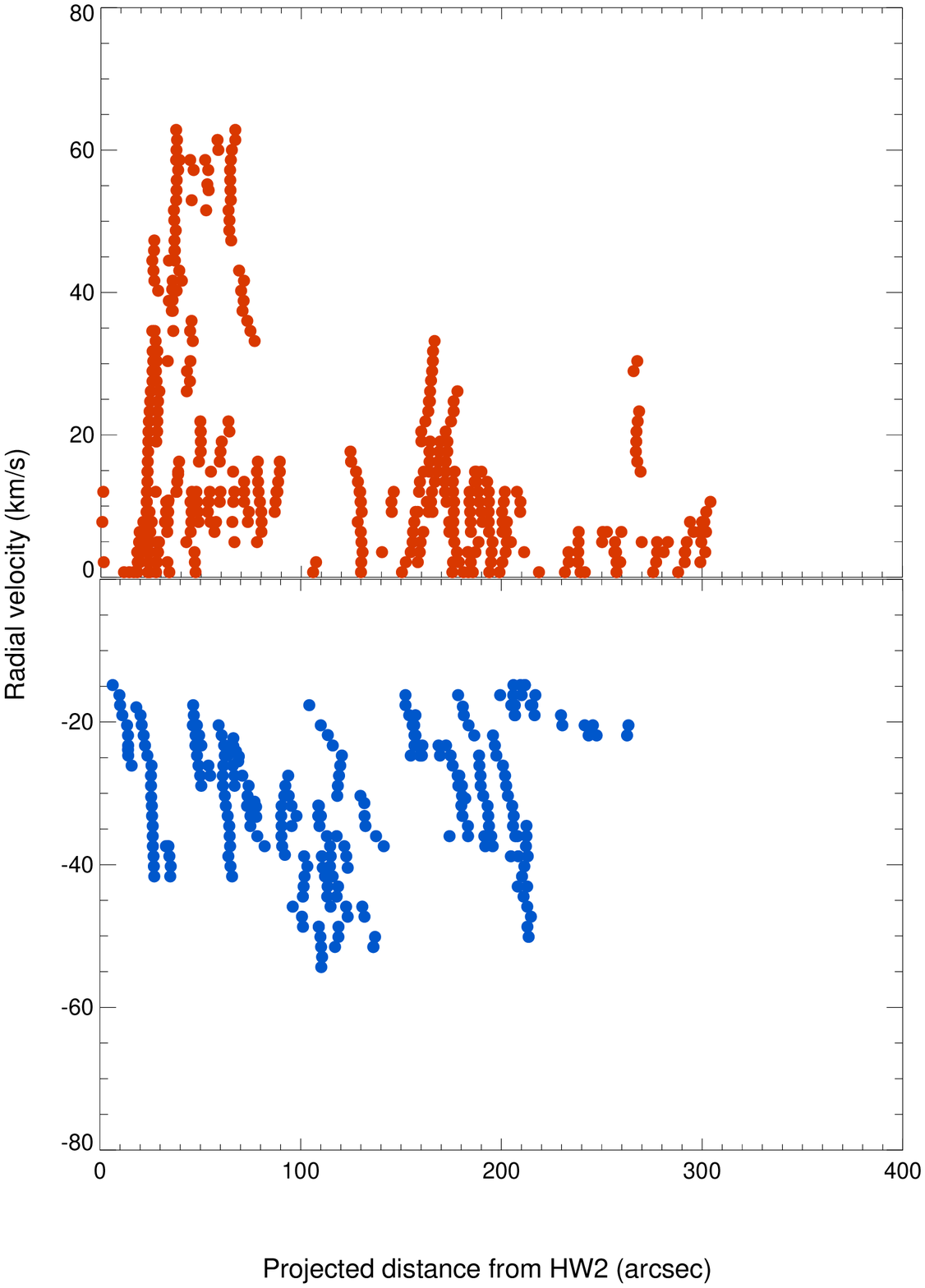}
\caption{\scriptsize Position-velocity relation of the $^{12}$CO(2-1)
structures from HW2: radial velocity as a function of on-the-sky
distance from the position of HW2. Note that velocities between $-$13
and 0 km s$^{-1}$ could not be investigated because of interferometric
contamination with the extended molecular gas. .}
\label{fig7}
\end{center}
\end{figure}

In Figure 7, we show a similar digram to that made in Figure 6, but now plotting 
the average position angles {\it vs.} the average radial velocities of the $^{12}$CO
compact condensations.  These average radial velocities are obtained averaging 
the radial velocities of each condensation that are inside of the dashed box
presented in Figure 6. In this way, for one average angle in the dashed box,
we have a average radial velocity.  This plot reveals that the recent ejections of the
outflow in average have small position angles and small radial
velocities, suggesting that the outflow is thus approaching to the
plane of the sky at present. The past ejections in average have large position
angles and large radial velocities indicating that the outflow in the near past was
probably located perpendicular to the plane of the sky. However, the
condensations with the largest radial velocities also show low radial
velocities. This suggests that maybe long time ago the outflow was
ejecting material close to the plane of the sky as it is doing now. 

Our plots revealed only five ejections separated by about 10$^\circ$, one more than
\citet{cun2009}. As mentioned earlier, from our data it is not clear
that the object HH 174 was ejected from HW2.  We found very
high velocity molecular gas associated with HW2 that is not presented
in our Figures 5 and 6, but either this is presented in our moment zero map (Figure 1).
For example, in our Figure 5, this emission represents a single multicolor
point in the {\it (0,0)} position that does not contribute to our findings. 
Perhaps this high velocity material is part of very new 
ejections of HW2 with high velocity. The present size and orientation of the 
molecular outflow is shown
in Figure 3.  The present orientation of the outflow coincides very well
with that of the powerful thermal jet \citep{cur2006,rod1994}. 

In Figure 8, we show the kinematics of the molecular gas associated
with the precessing outflow powered by HW2. This position-velocity
diagram reveals that the kinematics of the multiple flows is very
different of that observed in the explosive outflow in Orion BN/KL and DR21
\citep{zap2009,zap2011,zap2013}. The molecular condensations from the
precessing outflow mostly stay around some place with respect to a range of velocities 
and show large velocity gradients, contrary to the Hubble Law observed in the
molecular filaments arising from Orion BN/KL or DR21, see Figures 2 of
\citet{zap2009, zap2013}. In the case of Orion BN/KL and DR21, the molecular
condensations move in position when the velocity change, following a linear
correlation between velocity and position.
In conclusion, we suggest that an energetic explosive event as the
one occurred in Orion BN/KL does not explain the kinematics of the
multiple molecular outflows found here.   The kinematics of the molecular ejections 
observed in the CepA HW2/3 region is better explained by the 
presence of a precesing outflow.

From Figure 8, it is clear that the $^{12}$CO line emission with the largest redshifted 
line widths are located close to HW2 and from Figure 2 one can see that the it 
coincides with a high density ammonia core. We have marked this region with 
a circle in Figure 2. We think that this high density region might be blocking 
part of the redshifted side of the molecular outflow creating a ''shocked zone''.
 This "shocked zone" could explain why it is not observed the redshifted
bow shocks with similar position angles as the blueshifted bow shocks
if both sides come from the same precessing bipolar outflow. We refer again 
to the readers see \citet{chema1993} for the physical values of this ammonia 
condensation.

\subsubsection{The bipolar outflows from HW3c/d}

Our observations revealed the presence of two east-west low-velocity
bipolar outflows that are emanating from HW3c/d. The innermost parts of
the outflows appear to be traced by elongated ionized thermal jets
reported by \citet{garay1996,rod1994,rod2005} (see Figure 4). The
outflow powered by HW3d appears to be creating the zone called HW7 located to the
southeast and forming some more eastern HH objects far from HW3. 
The proper motions of HW7 are complex, however, \citet{rod2005} suggested that
they are being produced by a source to its northwest, perhaps by
HW3. This is in good agreement to what we propose here based 
on our $^{12}$CO observations.  On the other hand, the blueshifted side 
of the outflow powered by HW3c goes to the west and seems to connect with the
optical/X-Ray object, know as the HH 168 or GDD 37
\citep{pra2009,sch2009a,sch2009b,har2000,har1986}.  HH 168 is also
blueshifted \citep{Hi2004}. 

There is also the possibility that the
radio objects W1-4 \citep{rod2005} are also energized by the
blueshifted side of the outflow from HW3c, however, their radial
velocities are still unknown. These objects show westward proper
motions in the range of 120 to 280 $\kms$. \citet{rod2005} proposed
that these radio objects are energized by HW3.

\citet{chi2012} proposed that the object HW3d harbors an internal
massive young star that powers a NW-SE water maser outflow and the
thermal radio jet. This source could be energizing the object HW7
and perhaps the most eastern HH objects as mentioned before. 
However, we do not find any submillimeter continuum emission associated with this
source.  Perhaps this source is associated with a lower mass young stellar object 
and more sensitive millimeter observations are needed to reveal the 
dust emission associated to this protostar.

\section{Summary}

We have observed the Cepheus A East region using the Submillimeter
Array. We made (sub)millimeter line and continuum observations in a
mosaicing mode of this massive star forming region. Our mosaic covers
a total area of about 3$'$ $\times$ 12$'$, centered in the HW 2/3
region. Our main conclusions are as follows:

\begin{itemize}
\item We report compact and high velocity $^{12}$CO(2-1) emission
associated with the multiple east-west H$_2$ bow shocks present in 
Cepheus A East.   Blueshifted and redshifted gas emission is found
towards the east as well as to the west from HW2. Receding
$^{12}$CO features show radial velocities up to 65 $\kms$ 
while the approaching 12CO line emission has velocities down to $-$55 $\kms$.

\item The $^{12}$CO(2-1) observations suggest the presence of three large-scale 
east-west outflows, likely powered at small scales by the radio sources associated 
with the young massive protostars HW2, HW3c and HW3d.

\item We confirm the precession associated with the molecular outflow
powered by HW2.  Our data reveal five periodic ejections of material separated
approximately every 10$^\circ$.  There is some evidence that this outflow had an orientation
perpendicular to the plane of the sky on the past.

\item An energetic explosive event as the one that occurred in Orion BN/KL or DR21
does not explain the kinematics of the multiple molecular outflows
found here.

\item In our entire field, the continuum observations only revealed a
strong millimeter source associated with the HW 2/3 region.  We
resolved this extended dusty object in only two compact sources
associated with HW2 and HW3c.  These compact objects are
tracing the circumstellar disks associated with these objects.

\item The bright optical/X-Ray HH 168 object might be produced by
strong shocks with the
molecular cloud and arising from the outflow energized by HW3c.

\end{itemize}

In conclusion, the bipolar flows arising from HW2 and HW3c/d might
explain the whole outflow activity observed in Cepheus A East.
However, many more observations at different wavelengths are needed to
confirm this hypothesis. In particular, future observations using other molecular 
species may reveal more outflows in this high-mass forming region.

\acknowledgments

L.A.Z., L.F.R. and S.C. acknowledge the financial support from DGAPA,
UNAM, and CONACyT, M\'exico.  We are very grateful to John Bally for
having provided the H$_2$ image.

\end{document}